\documentclass[seceq,twocolumn]{jpsj2} %% two-column layout
\def\runauthor{Hiroaki \textsc{Kusunose} and Hiroaki \textsc{Ikeda}}

\title{
Interplay of crystal field structures with $f^2$ configuration to heavy fermions
}

\author{
Hiroaki \textsc{Kusunose}$^{1}$\thanks{kusu@cmpt.phys.tohoku.ac.jp}
and
Hiroaki \textsc{Ikeda}$^{2}$
}

\inst{
$^{1}$Department of Physics, Tohoku University, Sendai 980-8578 \\
$^{2}$Department of Physics, Kyoto University, Kyoto 606-8502
}

\abst{
We examine a relevance between characteristic of crystal field structures and heavily renormalized quasiparticle states in the $f^0$-$f^1$-$f^2$ Anderson lattice model.
Using a slave-boson mean-field approximation, we find that for $f^2$ configurations two or three quasiparticle bands are formed near the Fermi level depending on the number of the relevant $f^1$ orbitals in the $f^2$ crystal field {\it ground state}.
The inter-orbital correlations characterizing the crystal field ground state closely reflect in inter-band residual interactions among quasiparticles.
Particularly in the case of a singlet crystal field ground state, resulting residual {\it antiferromagnetic} exchange interactions among the quasiparticles lead to an anomalous suppression of the quasiparticle contribution of the spin susceptibility, even though the quasiparticle mass is strongly enhanced.
}

\kword{
U-based heavy fermion, Pr-based filled skutterudite, crystalline electric field, slave boson
}

\begin{document}
\maketitle

\section{Introduction}
A variety of heavy-fermion behaviors has been observed so far in systems such as Ce-based, U-based compounds, Pr-based filled skutterudites, etc., some of which exhibits an unconventional superconductivity and/or a peculiar phase transition\cite{Hewson93,Kuramoto00,Sugawara02,Maple03}.
Since an $f$-electron valency is usually within $f^0$-$f^1$ in the Ce-based systems, it is meaningful to investigate a non-degenerate Anderson lattice model, where the preoccupied orbital among all $f$ orbitals is retained.
For such systems a Landau quasiparticle (QP) approach\cite{Yamada86,Rice85,Millis87,Auerbach86,Newns87} is successful both conceptually and {\it practically} because of two reasons: (i) the relevant crystalline electric field (CEF) multiplet with $f^1$ configuration is considered essentially in a single-particle picture with the Kramers degeneracy, which turns into a dominant component of heavy-mass QP band, (ii) a residual interaction among QPs could be deduced from a {\it single} parameter, i.e., a renormalized on-site repulsion, which is of the order of the renormalized bandwidth.

On the contrary, the U-based systems and the Pr-based skutterudites are considered to have the $f$-electron valency close to 2. In this case, the $f^2$ CEF multiplets are described not only by filling of the $f^1$ CEF orbitals, but also taking into account intra-atomic interactions like inter-orbital repulsions, Hund's coupling and so on\cite{Hutchings65,Newns87}.
Moreover the $f^2$ CEF multiplets generally lack the Kramers degeneracy.
Although the QP approach may still be useful conceptually for a description of these heavy fermions, a relevance of the CEF structures to a nature of the QP and its residual interactions are non-trivial problem in practice.
A simple perturbational treatment does not work since a characteristic of the $f^2$ CEF multiplets would be lost if the intra-atomic interactions are treated in few lowest orders.

The purpose of this paper is to understand what influence the $f^2$ CEF multiplet correlations have on a feature of the QPs.
We examine the ground state (GS) of the Anderson lattice model using a slave-boson mean-field approach\cite{Kotliar86,Li89,Trees95}, in which the local Hilbert space is restricted to $f^0$-$f^1$-$f^2$ configurations.
It is the simplest way to discuss a renormalized QP band and it has the same physical ground of the Gutzwiller approximation\cite{Kotliar86}.
Trees {\it et. al.} first discussed a Fermi liquid description in the $f^1$-$f^2$ Anderson lattice model using a simplified CEF states for simplicity\cite{Trees95}.
Ikeda and Miyake suggested that a new type of Fermi liquid state is realized for the CEF singlet GS\cite{Ikeda97}.

In this paper we demonstrate a close relevance of the characteristic of the $f^2$ CEF {\it ground state} to a nature of QP bands.
In accordance with a proper symmetry property of the $f^2$ CEF states, we demonstrate that three QP bands appear near the Fermi level for an $f^2$ CEF GS that is made of three $f^1$ orbitals.
As a special case that only two $f^1$ orbitals are concerned in the $f^2$ CEF GS, two QP bands are formed near the Fermi level as was shown in the previous studies\cite{Trees95,Ikeda97}.
In the case of a singlet CEF GS, the intra-atomic interactions turn into inter-band {\it antiferromagnetic} exchange interactions, which act to quench spin degrees of freedom of the QPs.
It is a natural consequence to gain a local correlation energy of the CEF state as well as a coherent kinetic energy of the QPs.

The paper is organized as follows. In the next section (\S2), we introduce a model and slave-boson mean-field equations for general $f^0$-$f^1$-$f^2$ Anderson lattice model.
In \S3, we present the QP density of states (DOS), the renormalization factors and the renormalized $f$ levels as a function of the total $f$ electron numbers in the case of two different CEF GSs under the hexagonal symmetry as an example.
The difference of the QP contribution to the magnetic susceptibility is demonstrated for two CEF GSs.
The final section ($\S4$) summarizes the paper.
The appendix A is given for a comprehensive list of the $f^2$ CEF states for cubic, hexagonal and tetragonal symmetries, which are expressed in terms of the direct products of the $f^1$ CEF states.
The appendix B is given for necessary formulae of the mean-field calculation.

\section{Model and formulation}
The Hamiltonian for the $f^0$-$f^1$-$f^2$ Anderson lattice model is given by
\begin{equation}
H=\sum_{{\mib k}\gamma\sigma}\epsilon_{{\mib k}}c^\dagger_{{\mib k}\gamma\sigma}c_{{\mib k}\gamma\sigma}+V\sum_{{\mib k}\gamma\sigma}(f^\dagger_{{\mib k}\gamma\sigma}c_{{\mib k}\gamma\sigma}+{\rm h.c.})+H_f,
\label{hamiltonian}
\end{equation}
where we have used ${\mib k}$ and orbital independent hybridizations between the conduction and $f$ electrons with the same irreducible representation $\Gamma_\gamma$ of the point group ($\sigma$ is the pseudo spin for the Kramers pairs), and $\gamma$ independent dispersion, $\epsilon_{\mib k}$ for simplicity.
The localized Hamiltonian $H_f=\sum_iH_f^{(i)}$ in the $f^0$-$f^1$-$f^2$ restricted Hilbert space is given by
\begin{multline}
H_f^{(i)}=\sum_{\gamma\sigma}(E_f(1)+\Delta E_{1\gamma})|\Gamma_\gamma\sigma\rangle_i{}_i\langle\Gamma_\gamma\sigma| \\
+\sum_{\mu\nu}(E_f(2)+\Delta E_{2\mu})|\Gamma_\mu\nu\rangle_i{}_i\langle\Gamma_\mu\nu|,
\end{multline}
where $E_f(n_f)=E_fn_f+Un_f(n_f-1)/2$ is the $f$ electron energy for the total $f$ number, $n_f$, without the CEF splitting.
The energy splitting is introduced by $\Delta E_{n\gamma}$ for the $f^n$ $\Gamma_\gamma$ CEF states.

Let us introduce the auxiliary bosons, $e_i$, $p_{i\gamma\sigma}$ and $d_{i\mu\nu}$.
The squares of the classical values of these fields are supposed to give the occupation probabilities of the $f^0$, $f^1$  and $f^2$ CEF states respectively.
The introduction of the bosons allows us to linearize the interaction terms implicitly appeared in $H_f^{(i)}$ and to eliminate the fermion degrees of freedom.
The resultant system of interacting bosons may be shown to have a paramagnetic mean-field solution (we omit the site indices, $i$, for mean-field values), which corresponds to a description of the Fermi liquid.

The mean-field bosons must satisfy the completeness relation,
\begin{equation}
I=e^2+\sum_{\gamma\sigma}p_{\gamma\sigma}^2+\sum_{\mu\nu}d_{\mu\nu}^2=1.
\label{completeness}
\end{equation}
Two different counting for the $f$-electron number should coincide, namely,
\begin{equation}
Q_{\gamma\sigma}=p_{\gamma\sigma}^2+\sum_{\mu\nu}|\langle \Gamma_\gamma\sigma|\Gamma_\mu\nu\rangle|^2d_{\mu\nu}^2=f^\dagger_{i\gamma\sigma}f_{i\gamma\sigma},
\label{fnumber}
\end{equation}
where the coefficient in the second term represents the probability of the $f^1$ $|\Gamma_\gamma\sigma\rangle$ state in the $f^2$ $|\Gamma_\mu\nu\rangle$ state.
The explicit expressions of the CEF states for cubic, tetragonal and hexagonal symmetries are summarized in Appendix A.
In order to describe the hopping process of bosons accompanying any hopping process of $f$ electrons, we also introduce projection operators, $z_{\gamma\sigma}$, i.e., $f^\dagger_{{\mib k}\gamma\sigma}c_{{\mib k}\gamma\sigma}\to z_{\gamma\sigma}f^\dagger_{{\mib k}\gamma\sigma}c_{{\mib k}\gamma\sigma}$.
It is given by
\begin{equation}
z_{\gamma\sigma}=w_{\gamma\sigma}(p_{\gamma\sigma}e+\sum_{\mu\nu}|\langle \Gamma_\gamma\sigma|\Gamma_\mu\nu\rangle|^2d_{\mu\nu}p_{\gamma\sigma}).
\label{projection}
\end{equation}
Since the normalization of the projection operators is not unique, it is natural to choose $w_{\gamma\sigma}$ such that the total transition probability becomes unity, i.e.,
\begin{equation}
w_{\gamma\sigma}=Q_{\gamma\sigma}^{-1/2}(1-Q_{\gamma\sigma})^{-1/2}.
\end{equation}
Then, the conservation of the probability is ensured and the correct weak-coupling limit is guaranteed.
With this prescription the mean-field solution is known to be identical to Gutzwiller's solution\cite{Kotliar86,Li89,Trees95}.

Finally, we obtain the mean-field Hamiltonian which can be diagonalized as,
\begin{equation}
H_{\rm MF}=\sum_{{\mib k}m\gamma\sigma} E_{{\mib k}\gamma\sigma}^{(m)}a^\dagger_{{\mib k}m\gamma\sigma}a_{{\mib k}m\gamma\sigma}+NH_{\rm b},
\end{equation}
where $N$ is the number of sites.
The hybridized QP bands are given by
\begin{equation}
E_{{\mib k}\gamma\sigma}^{(\pm)}=\frac{1}{2}\biggl[
\epsilon_{\mib k}+\lambda_{\gamma\sigma}\pm\sqrt{(\epsilon_{\mib k}-\lambda_{\gamma\sigma})^2+4q_{\gamma\sigma}V^2}\biggr],
\end{equation}
where the Lagrange multiplier, $\lambda_{\gamma\sigma}$, for the $f$-number constraint, eq.~(\ref{fnumber}), acts as the renormalized $f$ level, and $q_{\gamma\sigma}=z_{\gamma\sigma}^2$ is the renormalization factor for the ($\Gamma_\gamma\sigma$) band.
The boson terms are given by
\begin{multline}
H_{\rm b}=\sum_{\gamma\sigma}(E_f(1)+\Delta E_{1\gamma})p_{\gamma\sigma}^2+\sum_{\mu\nu}(E_f(2)+\Delta E_{2\mu})d_{\mu\nu}^2 \\
+\lambda(I-1)-\sum_{\gamma\sigma}\lambda_{\gamma\sigma}Q_{\gamma\sigma},
\end{multline}
where additional Lagrange multiplier, $\lambda$, appears to satisfy eq.~(\ref{completeness}).

Minimizing the GS energy with respect to the bosons and the Lagrange multipliers, we obtain the set of the mean-field equations,
\begin{subequations}
\begin{align}
&2V\sum_{\gamma\sigma}\frac{\partial z_{\gamma\sigma}}{\partial{\rm b}}W_{\gamma\sigma}+\frac{\partial H_{\rm b}}{\partial{\rm b}}=0,\,\,\,({\rm b}=e,\;p_{\gamma\sigma},\;d_{\mu\nu}) \\
&I=1,\\
&Q_{\gamma\sigma}=n_{f\gamma\sigma},\\
&n=\sum_{\gamma\sigma}(n_{c\gamma\sigma}+n_{f\gamma\sigma}),
\end{align}
\end{subequations}
where the last equation is to determine the chemical potential $\mu$ for given total electron density, $n$.
Here we have defined the expectation values as
\begin{subequations}
\begin{align}
&W_{\gamma\sigma}=\frac{1}{N}\sum_{\mib k}\langle f^\dagger_{{\mib k}\gamma\sigma}c_{{\mib k}\gamma\sigma}\rangle, \label{evalue1} \\
&n_{c\gamma\sigma}=\frac{1}{N}\sum_{\mib k}\langle c^\dagger_{{\mib k}\gamma\sigma}c_{{\mib k}\gamma\sigma}\rangle, \label{evalue2} \\
&n_{f\gamma\sigma}=\frac{1}{N}\sum_{\mib k}\langle f^\dagger_{{\mib k}\gamma\sigma}f_{{\mib k}\gamma\sigma}\rangle, \label{evalue3}
\end{align}
\end{subequations}
whose expressions can be obtained analytically by using the rectangular DOS of conduction electrons (see Appendix B).
Note that in the absence of the magnetic field, we can omit the $\sigma$ and $\nu$ dependences in the mean-field equations.

\section{Results}
In this section we discuss a relevance of the CEF structure to a nature of the QP bands taking the hexagonal symmetry as an example. It is emphasized that the close connection between the CEF GS and the resultant QPs shown in this section is hold in other symmetries as well.
For simplicity, we retain only the two lowest CEFs in the $f^2$ configuration, {\it e.g.}, the $|\Gamma_4\rangle$ singlet and the $|\Gamma_5^{(1)}\pm\rangle$ doublet with $\alpha^2=\beta^2=1/2$.
The necessary quantities are given as follows:
\begin{equation}
I=e^2+\sum_{\gamma\sigma}p_{\gamma\sigma}^2+d_4^2+\sum_\nu d_{5\nu}^2,
\end{equation}
\begin{subequations}
\begin{align}
& Q_{7\sigma}=p_{7\sigma}^2+\frac{1}{2}d_4^2+\frac{\beta^2}{14}(5d_{5\sigma}^2+9d_{5-\sigma}^2), \\
& Q_{8\sigma}=p_{8\sigma}^2+\frac{1}{2}d_4^2+(\alpha^2+\frac{9\beta^2}{14})d_{5\sigma}^2, \\
& Q_{9\sigma}=p_{9\sigma}^2+(\alpha^2+\frac{5\beta^2}{14})d_{5\sigma}^2,
\end{align}
\end{subequations}
\begin{subequations}
\begin{align}
& z_{7\sigma}=p_{7\sigma}e+\frac{1}{2}d_4p_{8\sigma}+\frac{\beta^2}{14}(5d_{5\sigma}p_{9\sigma}+9d_{5-\sigma}p_{8-\sigma}), \\
& z_{8\sigma}=p_{8\sigma}e+\frac{1}{2}d_4p_{7\sigma}+d_{5\sigma}(\alpha^2p_{9\sigma}+\frac{9\beta^2}{14}p_{7-\sigma}), \\
& z_{9\sigma}=p_{9\sigma}e+d_{5\sigma}(\alpha^2p_{8\sigma}+\frac{5\beta^2}{14}p_{7\sigma}).
\label{hexz9}
\end{align}
\end{subequations}
We use the constant $f^1$ CEF splitting, $\Delta E_7=-0.1$, $\Delta E_8=0$, $\Delta E_9=0.1$ and the Coulomb repulsion, $U=3$ and the hybridization strength, $V=0.2$ in the unit of $D=1$ throughout this paper. The $f^2$ CEF splitting is parametrized by $\Delta$ as $\Delta E_4=-2\Delta$ and $\Delta E_5=\Delta$.
We fix the total electron density, $n=4$ and vary the average $f$ level, $E_f$, to adjust the total $f$-electron number, $n_f=\sum_{\gamma\sigma}n_{f\gamma\sigma}$.

First, we consider the case of $\Delta=0.1$, i.e., the $\Gamma_4$ GS.
\begin{figure}
\includegraphics[width=8.5cm]{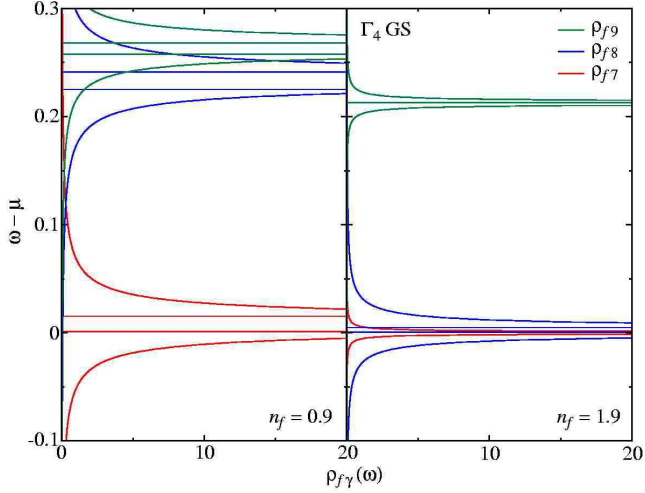}
\caption{The quasiparticle (QP) density of states (DOS) for $\Gamma_4$ $f^2$ CEF ground state (GS). Reflecting the $f^1$ CEF ground state, $\Gamma_7$ QP band is formed for $n_f\sim1$. For $n_f\sim2$ two relevant QP bands ($\Gamma_7$ and $\Gamma_8$) develop near the chemical potential.}
\label{g4-dos}
\end{figure}
Figure~\ref{g4-dos} shows the QP DOS in the $\Gamma_\gamma$ band, $\rho_{f\gamma}(\omega)$.
Reflecting the fact that the $\Gamma_7$ CEF is the GS in $f^1$ configuration, the $\Gamma_7$ QP band develops near the chemical potential for $n_f\sim 1$.
This indicates the validity of a description for $f^1$-based systems using the non-degenerate Anderson lattice model.
The rest of two QP bands appear far from the chemical potential.
With further decrease of $E_f$, $n_f\sim2$ is realized, in which two QP bands ($\Gamma_7$ and $\Gamma_8$) appear near $\mu$ leaving the $\Gamma_9$ band in the high-energy region.
As will be shown in the comparison with the case of $\Gamma_5^{(1)}$ GS, this is a consequence that the $\Gamma_4$ GS only involves two $f^1$ states (see eq.~(\ref{hexg4})).

\begin{figure}[t]
\includegraphics[width=8.5cm]{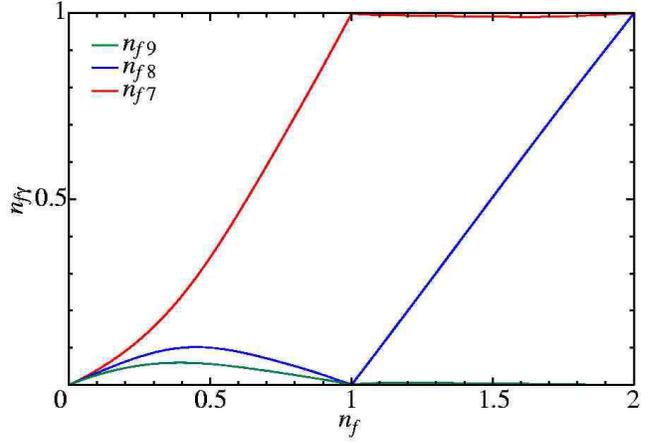}
\caption{The $f$-electron density in the $\gamma$ band as a function of the total $f$ electron density for $\Gamma_4$ $f^2$ CEF GS.}
\label{g4-nf}
\end{figure}
The band distribution of the $f$-electron occupation is shown in Fig.~\ref{g4-nf}.
For $n_f<1$ the band with the lowest $f^1$ CEF state (i.e. $\Gamma_7$) is first filled, while for $1<n_f<2$ the bands connecting to the $f^2$ CEF GS (i.e. $\Gamma_7$ and $\Gamma_8$) are occupied.

\begin{figure}
\includegraphics[width=8.5cm]{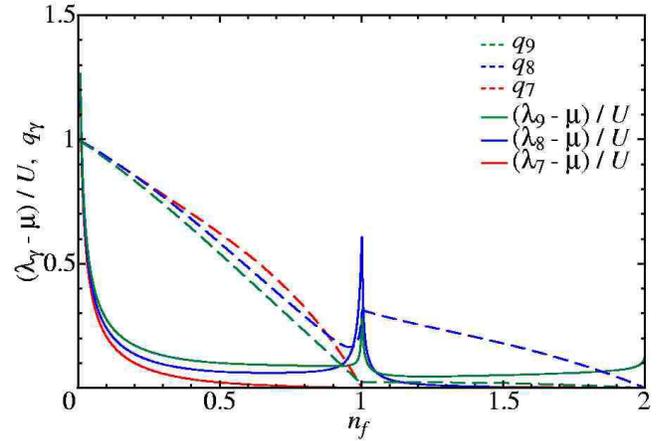}
\caption{The renormalized $f$ level measured from the chemical potential and the renormalization factor for $\Gamma_4$ $f^2$ CEF GS.}
\label{g4-e-q}
\end{figure}
In Fig.~\ref{g4-e-q} the renormalized $f$ level, $\lambda_\gamma$, and the renormalization factor, $q_\gamma$, are shown as a function of $n_f$.
The two of the three $f$ levels are renormalized to locate near $\mu$ leaving the rest in high-energy region toward $n_f=2$.
This vanishing CEF splitting between two renormalized bands has already been recognized\cite{Trees95,Ikeda97}.
In the limit $n_f\to2$, the renormalization factor $q_9$ becomes very small.
This is due to an empty of the $\Gamma_9$ band, namely, $e\sim p_9\sim d_5\sim 0$ (see eq.~(\ref{hexz9})), and is regarded as an artifact of the slave-boson mean-field approximation.

At this point, we should mention drawback of the slave-boson mean-field approximation.
At the mean-field level all electronic states must necessarily be expressed as itinerant coherent bands\cite{Raimondi93}.
In reality, only the states near the chemical potential are coherent in strongly correlated systems.
Therefore the appearance of the highly renormalized QP band far from the chemical potential is unreliable, and may give a measure of the incoherent CEF excitations at the very most.
A criterion of reliable result is whether the renormalized QP bands appear in the vicinity of the chemical potential.

Next, we consider the case of the $\Gamma_5^{(1)}$ GS ($\Delta=-0.1$).
\begin{figure}
\includegraphics[width=8.5cm]{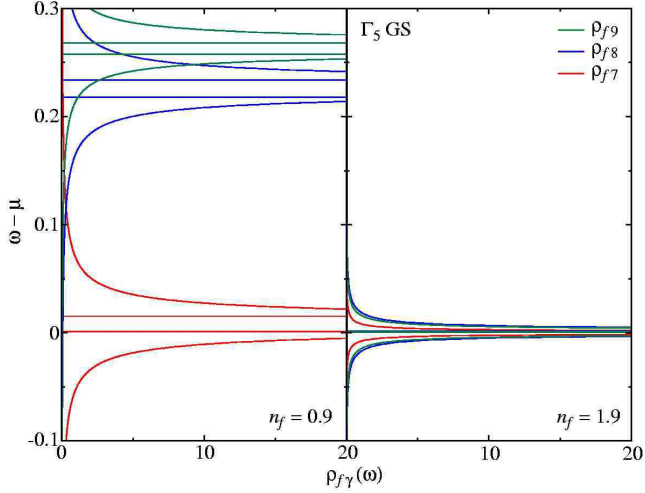}
\caption{The QP DOS for $\Gamma_5^{(1)}$ $f^2$ CEF GS. For $n_f\sim2$ all the three QP bands are formed reflecting $\Gamma_5^{(1)}$ GS.}
\label{g5-dos}
\end{figure}
The QP DOS for $n_f\sim 1$ and $n_f\sim2$ are shown in Fig.~\ref{g5-dos}.
The former case is roughly the same as that for the $\Gamma_4$ GS.
This again indicates the usefulness of the $f^0$-$f^1$ non-degenerate Anderson lattice model in discussing $f^1$-based systems.
In contrast to the $\Gamma_4$ GS, three QP bands are formed near $\mu$ in the case of $n_f\sim2$. 
This is a reflection of the $\Gamma_5^{(1)}$ GS that is made of all $f^1$ CEF states (see eq.(\ref{hexg5})).

\begin{figure}
\includegraphics[width=8.5cm]{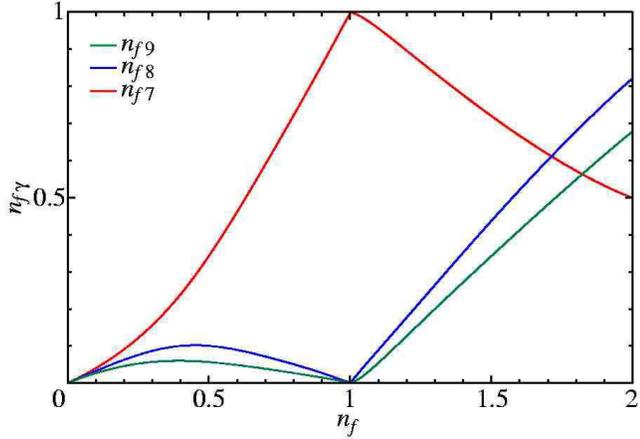}
\caption{The $f$-electron density distribution to each QP bands for the $\Gamma_5^{(1)}$ GS.}
\label{g5-nf}
\end{figure}
The band distribution of the $f$-electron occupation is shown in Fig.~\ref{g5-nf}.
The behavior of the $f$-electron distribution for $n_f<1$ is almost equal to that in the $\Gamma_4$ GS.
On the contrary the tendency of the distribution is completely different for $n_f>1$; it goes toward the ratio of $n_{f7}:n_{f8}:n_{f9}=0.5:0.82:0.62$.
This ratio is nothing but the probability of each $f^1$ states in the $\Gamma_5^{(1)}$ state.
It is a natural consequence to gain a local correlation energy of the CEF state as well as a coherent kinetic energy of the QPs.
From this point of view, the result for the $\Gamma_4$ GS is understood as the special case of $n_{f7}:n_{f8}:n_{f9}=1:1:0$.

\begin{figure}
\includegraphics[width=8.5cm]{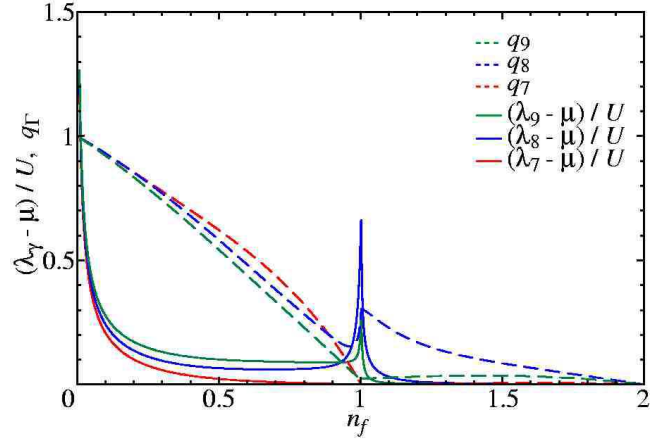}
\caption{The renormalized $f$ level and the renormalization factor for the $\Gamma_5^{(1)}$ GS.}
\label{g5-e-q}
\end{figure}
Figure~\ref{g5-e-q} confirms that both the effective $f$ levels and the bandwidths for all the QP bands are highly renormalized as $n_f$ approaches to 2.

Lastly, we investigate the QP part of the uniform magnetic susceptibility, $\chi_{\rm QP}$, which is obtained by the numerical differentiation, $\chi_{\rm QP}=\partial \sum_{\gamma\sigma}g_\gamma\sigma(n_{c\gamma\sigma}+n_{f\gamma\sigma})/\partial h$ with applied the tiny magnetic field ($h=10^{-6}$).
The $g_\gamma$ factor is given by $g_\gamma=2\langle{\Gamma_\gamma+}|j_z|\Gamma_\gamma+\rangle$.
Note that the slave-boson mean-field approximation cannot describe incoherent part of the single-particle spectral weight properly.
As a result, the susceptibility lacks contributions from those concerning the incoherent states, such as the Van Vleck susceptibility.
The susceptibility is thus regarded as the pure QP contributions from both the intra- and the inter-QP bands particle-hole excitations\cite{Nakano91}.
Since it is known to have an unphysical instability, i.e., $\chi_{\rm QP}<0$, in the Gutzwiller approximation\cite{Rice85}, we use no Gutzwiller generalization, i.e., $w_{\gamma\sigma}=1$, in calculating the QP susceptibility.

\begin{figure}
\includegraphics[width=8.5cm]{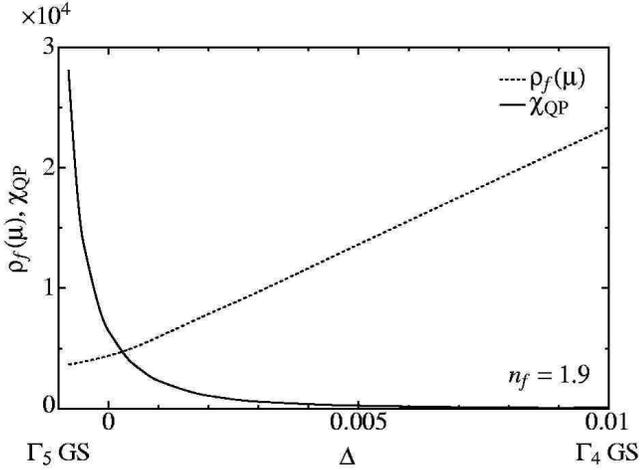}
\caption{The renormalized $f$-electron DOS and the uniform QP susceptibility as a function of the $f^2$ CEF splitting. The QP susceptibility is strongly suppressed, while the $f$-DOS remains strongly enhanced in the case of the $\Gamma_4$ GS.}
\label{g4g5-rf-chi}
\end{figure}
Figure~\ref{g4g5-rf-chi} shows the DOS at the Fermi energy and the QP susceptibility as a function of the $f^2$ CEF splitting, in which we have adjusted $E_f$ such that $n_f=1.9$ regardless of $\Delta$.
In the case of the $\Gamma_5$ GS, both $\rho_f(\mu)$ and $\chi_{\rm QP}$ are highly enhanced (of the order of $10^{4}$).
As the position of $\Gamma_4$ state decreases, $\chi_{\rm QP}$ is suppressed drastically, while $\rho_f(\mu)$ remains highly renormalized.
The similar result has been obtained by Ikeda and Miyake and interpreted by a magnetically inactive of the singlet CEF\cite{Ikeda97}.
We can understand this remarkable result in a consistent way of the case for $h=0$.
In the case of the $\Gamma_4$ GS at finite fields, the QP bands will develop satisfying the relation $n_{f+}=n_{f-}$, $n_{f\sigma}$ being $\sum_\gamma n_{f\gamma\sigma}$, in order to gain a local correlation energy of the CEF states (see eq.~(\ref{hexg4})) as much as possible.
Then, the QP exhibits the unrenormalized magnetic susceptibility, while the individual spin component of the QP bands are highly renormalized.
On the contrary, there is no such constraint on the ratio of $n_{f\sigma}$ for the $\Gamma_5^{(1)}$ doublet GS, giving rise to highly renormalized spin susceptibility.
This observation implies that the singlet correlation of the CEF GS sustains as strong interband {\it antiferromagnetic} exchange interactions to quench locally spin degrees of freedom of the QPs, which could be a source of a possible attractive interaction for unconventional superconductivities observed in various $f^2$-based heavy-fermion systems.
In the impurity analog of the present problem, the existence of such exchange interaction among QPs is confirmed for the singlet CEF GS by using Wilson's numerical renormalization-group calculations, \cite{Hattori}.

\section{Summary}
We have examined a relevance between characteristic of crystal field structures and heavily renormalized QP states in the $f^0$-$f^1$-$f^2$ Anderson lattice model.
Using a slave-boson mean field approximation, we have demonstrated a close connection of the nature of the $f^2$ CEF ground state with the peculiar character of the heavy fermions.
It is found that the plural QP bands appear near the Fermi level to maintain the ratio of the $f$-electron density in each bands being equal to that in the $f^2$ CEF ground state.
Such a formation of the QP bands is the optimization to gain a local correlation energy of the $f^2$ CEF state as well as a coherent kinetic energy of the QPs.
In the case of a singlet CEF ground state, this energetics requires the equal number of $f$ electrons for different spin components even at finite magnetic fields, leading to an anomalous suppression of the QP susceptibility with highly renormalized QP mass.
In the language of the Fermi liquid theory, there must exist strong interband {\it antiferromagnetic} exchange interactions among QPs to quench locally spin degrees of freedom, which could be a source of a possible attractive interaction for unconventional superconductivities observed in various $f^2$-based heavy-fermion systems.

\section*{Acknowledgment}
The authors would like to thank Y. Kuramoto, K. Miyake and H. Tou for stimulating discussions.
This work was supported by a Grant-in-Aid for Scientific Research Priority Area ``Skutterudite" of the Ministry of Education, Culture, Sports, Science and Technology, Japan.

\appendix
\section{The $f^2$ CEF states as a superposition of the $f^1$ CEF states}
Here we describe the $f^2$ CEF states as a linear combination of direct products of the $f^1$ CEF states.
This treatment for the $f^2$ CEF states is a point-group analog of the $j$-$j$ coupling scheme.

The $f^1$ CEF state, $|\Gamma_\gamma\sigma\rangle$ ($\gamma$ denotes an irreducible representation with a Kramers pair suffix $\sigma$), is written in terms of $|m\rangle$, the eigenstate of $j=5/2$, $j_z$,
\begin{equation}
|\Gamma_\gamma\sigma\rangle = \sum_m|m\rangle\langle m|\Gamma_\gamma\sigma\rangle.
\label{f1cef}
\end{equation}
Similarly the $f^2$ CEF state, $|\Gamma_\mu\nu\rangle$, is expressed by using $|M\rangle$, the eigenstate of $J=4$, $J_z$ as
\begin{equation}
|\Gamma_\mu\nu\rangle = \sum_M |M\rangle\langle M|\Gamma_\mu\nu\rangle.
\end{equation}
Using the relation in the $j$-$j$ coupling scheme,
\begin{equation}
|M\rangle=\sum_m |m\rangle|M-m\rangle\langle m;M-m|M\rangle,
\label{fpc}
\end{equation}
and the inverse of eq.~(\ref{f1cef}), we obtain a desired expression,
\begin{equation}
|\Gamma_\mu\nu\rangle = \sum_{\gamma\sigma\gamma'\sigma'}|\Gamma_\gamma\sigma\rangle|\Gamma_{\gamma'}\sigma'\rangle\langle \Gamma_\gamma\sigma;\Gamma_{\gamma'}\sigma'|\Gamma_\mu\nu\rangle,
\end{equation}
where the linear coefficients can be calculated by
\begin{multline}
\langle \Gamma_\gamma\sigma;\Gamma_{\gamma'}\sigma'|\Gamma_\mu\nu\rangle=\sum_{mM}\langle \Gamma_\gamma\sigma|m\rangle\langle\Gamma_{\gamma'}\sigma'|M-m\rangle
\times\\ \times
\langle m;M-m|M\rangle \langle M|\Gamma_\mu\nu\rangle.
\end{multline}
The probability in eqs.~(\ref{fnumber}) and (\ref{projection}) is then given by
\begin{equation}
|\langle\Gamma_\gamma\sigma|\Gamma_\mu\nu\rangle|^2=\sum_{\gamma'\sigma'}|\langle\Gamma_\gamma\sigma;\Gamma_{\gamma'}\sigma'|\Gamma_\mu\nu\rangle|^2.
\end{equation}
The explicit expressions for cubic, tetragonal and hexagonal symmetries are summarized below.

\subsection{cubic symmetry}
The definition of the $f^1$ CEF states is
\begin{subequations}
\begin{align}
&|\Gamma_7\pm\rangle=\sqrt{\frac{1}{6}}\biggl|\pm\frac{5}{2}\biggr\rangle-\sqrt{\frac{5}{6}}\biggl|\mp\frac{3}{2}\biggr\rangle, \\
&|\Gamma_{8a}\pm\rangle=\sqrt{\frac{5}{6}}\biggl|\pm\frac{5}{2}\biggr\rangle+\sqrt{\frac{1}{6}}\biggl|\mp\frac{3}{2}\biggr\rangle, \\
&|\Gamma_{8b}\pm\rangle=\biggl|\pm\frac{1}{2}\biggr\rangle.
\end{align}
\end{subequations}
The $f^2$ CEF states and their $j$-$j$ coupling expressions are given by
\begin{subequations}
\begin{align}
&|\Gamma_1\rangle=\frac{\sqrt{30}}{12}\biggl[|+4\rangle+|-4\rangle\biggr]+\frac{\sqrt{21}}{6}|0\rangle \nonumber \\
&\mbox{\hspace{0.6cm}}=\frac{1}{\sqrt{6}}\biggl[|0d0\rangle+|00d\rangle-2|d00\rangle\biggr], \\
&|\Gamma_3+\rangle=\frac{\sqrt{42}}{12}\biggl[|+4\rangle+|-4\rangle\biggr]-\frac{\sqrt{15}}{6}|0\rangle
\nonumber \\
&\mbox{\hspace{0.9cm}}=-\frac{2\sqrt{42}}{21}\biggl[|+-0\rangle-|-+0\rangle\biggr]
\nonumber\\
&\mbox{\hspace{1.3cm}}+\frac{\sqrt{210}}{42}\biggl[|0d0\rangle-|00d\rangle\biggr], \\
&|\Gamma_3-\rangle=\frac{1}{\sqrt{2}}\biggl[|+2\rangle+|-2\rangle\biggr]
\nonumber \\
&\mbox{\hspace{0.9cm}}=\frac{2\sqrt{42}}{21}\biggl[|+0-\rangle-|-0+\rangle\biggr]
\nonumber\\
&\mbox{\hspace{1.3cm}}+\frac{\sqrt{210}}{42}\biggl[|0+-\rangle-|0-+\rangle\biggr], \\
&|\Gamma_4\pm\rangle=\mp\biggl[\sqrt{\frac{1}{8}}|\mp3\rangle+\sqrt{\frac{7}{8}}|\pm1\rangle\biggr]
\nonumber\\
&\mbox{\hspace{0.9cm}}=-\frac{1}{2}|\pm\pm0\rangle+\frac{\sqrt{3}}{2}|\mp0\mp\rangle, \\
&|\Gamma_40\rangle=\frac{1}{\sqrt{2}}\biggl[|+4\rangle-|-4\rangle\biggr]
\nonumber \\
&\mbox{\hspace{0.8cm}}=\frac{1}{\sqrt{2}}\biggl[|+-0\rangle+|-+0\rangle\biggr], \\
&|\Gamma_5\pm\rangle=\pm\biggl[\sqrt{\frac{7}{8}}|\pm3\rangle-\sqrt{\frac{1}{8}}|\mp1\rangle\biggr]
\nonumber \\
&\mbox{\hspace{0.9cm}}=\frac{\sqrt{7}}{14}|\mp\mp0\rangle+\frac{\sqrt{21}}{42}|\pm0\pm\rangle+\frac{2\sqrt{105}}{21}|0\pm\pm\rangle, \\
&|\Gamma_50\rangle=\frac{1}{\sqrt{2}}\biggl[|+2\rangle-|-2\rangle\biggr]
\nonumber\\
&\mbox{\hspace{0.8cm}}=-\frac{1}{\sqrt{42}}\biggl[|+0-\rangle+|-0+\rangle\biggr]
\nonumber\\
&\mbox{\hspace{1.2cm}}+\frac{\sqrt{210}}{21}\biggl[|0+-\rangle+|0-+\rangle\biggr],
\end{align}
\end{subequations}
where $|\alpha\beta\gamma\rangle$ in the second equality represents the direct product of the $f^1$ CEF state, $|\Gamma_7\alpha\rangle|\Gamma_{8a}\beta\rangle|\Gamma_{8b}\gamma\rangle$, in which $d\equiv|\Gamma_\gamma+\rangle|\Gamma_\gamma-\rangle$ ($0$) indicates double (no) occupation.

\subsection{tetragonal symmetry}
The definition of the $f^1$ CEF states is
\begin{subequations}
\begin{align}
&|\Gamma_6\pm\rangle=\biggl|\pm\frac{1}{2}\biggr\rangle, \\
&|\Gamma_7^{(1)}\pm\rangle=\sqrt{\frac{5}{6}}\biggl|\pm\frac{5}{2}\biggr\rangle+\sqrt{\frac{1}{6}}\biggl|\mp\frac{3}{2}\biggr\rangle, \\
&|\Gamma_7^{(2)}\pm\rangle=\sqrt{\frac{1}{6}}\biggl|\pm\frac{5}{2}\biggr\rangle-\sqrt{\frac{5}{6}}\biggl|\mp\frac{3}{2}\biggr\rangle.
\end{align}
\end{subequations}
The $f^2$ CEF states are written as
\begin{subequations}
\begin{align}
&|\Gamma_1^{(1)}\rangle=\frac{\epsilon}{\sqrt{2}}\biggl[|+4\rangle+|-4\rangle\biggr]+\gamma|0\rangle \nonumber\\
&\mbox{\hspace{0.8cm}}=\biggl[\frac{\sqrt{10}}{6}\epsilon+\frac{\sqrt{14}}{42}\gamma\biggr]|0d0\rangle-\biggl[\frac{\sqrt{10}}{6}\epsilon+\frac{\sqrt{14}}{6}\gamma\biggr]|00d\rangle
\nonumber \\
&\mbox{\hspace{0.8cm}}+\sqrt{\frac{2}{7}}\gamma|d00\rangle-\biggl[\frac{\sqrt{2}}{3}\epsilon-\frac{\sqrt{70}}{21}\gamma\biggr]\biggl[|0+-\rangle-|0-+\rangle\biggr],\\
&|\Gamma_1^{(2)}\rangle=\frac{\gamma}{\sqrt{2}}\biggl[|+4\rangle+|-4\rangle\biggr]-\epsilon|0\rangle \nonumber\\
&\mbox{\hspace{0.8cm}}=\biggl[\frac{\sqrt{10}}{6}\gamma-\frac{\sqrt{14}}{42}\epsilon\biggr]|0d0\rangle-\biggl[\frac{\sqrt{10}}{6}\gamma-\frac{\sqrt{14}}{6}\epsilon\biggr]|00d\rangle
\nonumber \\
&\mbox{\hspace{0.8cm}}-\sqrt{\frac{2}{7}}\epsilon|d00\rangle-\biggl[\frac{\sqrt{2}}{3}\gamma+\frac{\sqrt{70}}{21}\epsilon\biggr]\biggl[|0+-\rangle-|0-+\rangle\biggr],\\
&|\Gamma_2\rangle=\frac{1}{\sqrt{2}}\biggl[|+4\rangle-|-4\rangle\biggr]\nonumber\\
&\mbox{\hspace{0.6cm}}=-\frac{1}{\sqrt{2}}\biggl[|0+-\rangle+|0-+\rangle\biggr], \\
&|\Gamma_3\rangle=\frac{1}{\sqrt{2}}\biggl[|+2\rangle+|-2\rangle\biggr]\nonumber\\
&\mbox{\hspace{0.6cm}}=\frac{2\sqrt{42}}{21}\biggl[|+0-\rangle-|-0+\rangle\biggr]\nonumber\\
&\mbox{\hspace{1cm}}+\frac{\sqrt{210}}{42}\biggl[|+-0\rangle-|-+0\rangle\biggr],\\
&|\Gamma_4\rangle=\frac{1}{\sqrt{2}}\biggl[|+2\rangle-|-2\rangle\biggr]\nonumber\\
&\mbox{\hspace{0.6cm}}=\frac{1}{\sqrt{42}}\biggl[|+0-\rangle+|-0+\rangle\biggr]\nonumber\\
&\mbox{\hspace{1cm}}-\frac{\sqrt{210}}{21}\biggl[|+-0\rangle+|-+0\rangle\biggr],\\
&|\Gamma_5^{(1)}\pm\rangle=\mp\biggl[\alpha|\mp3\rangle+\beta|\pm1\rangle\biggr]\nonumber\\
&\mbox{\hspace{1.1cm}}=-\biggl[\sqrt{\frac{5}{6}}\alpha-\sqrt{\frac{5}{42}}\beta\biggr]|\mp\mp0\rangle\nonumber\\
&\mbox{\hspace{1.5cm}}-\biggl[\sqrt{\frac{1}{6}}\alpha+\sqrt{\frac{25}{42}}\beta\biggr]|\mp0\mp\rangle+\sqrt{\frac{2}{7}}\beta|0\pm\pm\rangle,\\
&|\Gamma_5^{(2)}\pm\rangle=\pm\biggl[\beta|\pm3\rangle-\alpha|\mp1\rangle\biggr]\nonumber\\
&\mbox{\hspace{1.1cm}}=-\biggl[\sqrt{\frac{5}{6}}\beta+\sqrt{\frac{5}{42}}\alpha\biggr]|\pm\pm0\rangle\nonumber\\
&\mbox{\hspace{1.5cm}}-\biggl[\sqrt{\frac{1}{6}}\beta-\sqrt{\frac{25}{42}}\alpha\biggr]|\pm0\pm\rangle-\sqrt{\frac{2}{7}}\alpha|0\mp\mp\rangle,
\end{align}
\end{subequations}
where $\alpha^2+\beta^2=\gamma^2+\epsilon^2=1$ and $|\alpha\beta\gamma\rangle$ is an abbreviation of $|\Gamma_6\alpha\rangle|\Gamma_7^{(1)}\beta\rangle|\Gamma_7^{(2)}\gamma\rangle$.
Note that the cubic expressions are recovered when $\alpha=\sqrt{1/8}$, $\beta=\sqrt{7/8}$, $\gamma=\sqrt{21}/6$ and $\epsilon=\sqrt{15}/6$.

\subsection{hexagonal symmetry}
The definition of the $f^1$ CEF states is
\begin{subequations}
\begin{align}
&|\Gamma_7\pm\rangle=\biggl|\pm\frac{1}{2}\biggr\rangle, \\
&|\Gamma_8\pm\rangle=\biggl|\pm\frac{5}{2}\biggr\rangle, \\
&|\Gamma_9\pm\rangle=\biggl|\pm\frac{3}{2}\biggr\rangle.
\end{align}
\end{subequations}
The $f^2$ CEF states are given by
\begin{subequations}
\begin{align}
&|\Gamma_1\rangle=|0\rangle \nonumber\\
&\mbox{\hspace{0.6cm}}=\frac{1}{\sqrt{14}}\biggl[|0d0\rangle+3|00d\rangle+2|d00\rangle\biggr],\\
&|\Gamma_3\rangle=\frac{1}{\sqrt{2}}\biggl[|+3\rangle+|-3\rangle\biggr]\nonumber\\
&\mbox{\hspace{0.6cm}}=-\frac{1}{\sqrt{2}}\biggl[|++0\rangle-|--0\rangle\biggr],\\
&|\Gamma_4\rangle=\frac{1}{\sqrt{2}}\biggl[|+3\rangle-|-3\rangle\biggr]\nonumber\\
&\mbox{\hspace{0.6cm}}=-\frac{1}{\sqrt{2}}\biggl[|++0\rangle+|--0\rangle\biggr], \label{hexg4} \\
&|\Gamma_5^{(1)}\pm\rangle=\pm\biggl[\alpha|\pm4\rangle+\beta|\mp2\rangle\biggr] \nonumber\\
&\mbox{\hspace{1.1cm}}=\alpha|0\pm\pm\rangle+\beta\biggl[\sqrt{\frac{5}{14}}|\pm0\pm\rangle+\sqrt{\frac{9}{14}}|\mp\pm0\rangle\biggr], \label{hexg5} \\
&|\Gamma_5^{(2)}\pm\rangle=\pm\biggl[\beta|\pm4\rangle-\alpha|\mp2\rangle\biggr] \nonumber\\
&\mbox{\hspace{1.1cm}}=\beta|0\pm\pm\rangle-\alpha\biggl[\sqrt{\frac{5}{14}}|\pm0\pm\rangle+\sqrt{\frac{9}{14}}|\mp\pm0\rangle\biggr],\\
&|\Gamma_6\pm\rangle=\mp|\pm1\rangle\nonumber\\
&\mbox{\hspace{0.9cm}}=\sqrt{\frac{5}{7}}|\mp0\pm\rangle-\sqrt{\frac{2}{7}}|0\pm\mp\rangle,
\end{align}
\end{subequations}
where $\alpha^2+\beta^2=1$ and $|\alpha\beta\gamma\rangle$ is an abbreviation of $|\Gamma_7\alpha\rangle|\Gamma_8\beta\rangle|\Gamma_9\gamma\rangle$.

In deriving these expressions, it is useful to write the explicit forms of eq.~(\ref{fpc}) as
\begin{subequations}
\begin{align}
&|\pm4\rangle=\pm\biggl|\pm\frac{5}{2}\biggr\rangle\biggl|\pm\frac{3}{2}\biggr\rangle,\\
&|\pm3\rangle=\pm\biggl|\pm\frac{5}{2}\biggr\rangle\biggl|\pm\frac{1}{2}\biggr\rangle,\\
&|\pm2\rangle=\pm\biggl[\sqrt{\frac{5}{14}}\biggl|\pm\frac{3}{2}\biggr\rangle\biggl|\pm\frac{1}{2}\biggr\rangle+\sqrt{\frac{9}{14}}\biggl|\pm\frac{5}{2}\biggr\rangle\biggl|\mp\frac{1}{2}\biggr\rangle\biggr],\\
&|\pm1\rangle=\pm\biggl[\sqrt{\frac{5}{7}}\biggl|\pm\frac{3}{2}\biggr\rangle\biggl|\mp\frac{1}{2}\biggr\rangle+\sqrt{\frac{2}{7}}\biggl|\pm\frac{5}{2}\biggr\rangle\biggl|\mp\frac{3}{2}\biggr\rangle\biggr],\\
&|0\rangle=\frac{1}{\sqrt{14}}\biggl[\biggl|+\frac{5}{2}\biggr\rangle\biggl|-\frac{5}{2}\biggr\rangle+3\biggl|+\frac{3}{2}\biggr\rangle\biggl|-\frac{3}{2}\biggr\rangle+2\biggl|+\frac{1}{2}\biggr\rangle\biggl|-\frac{1}{2}\biggr\rangle\biggr].
\end{align}
\end{subequations}

\section{The form of the expectation values in eqs.~(\ref{evalue1})-(\ref{evalue3})}
In this appendix, we give explicit expressions for $W_{\gamma\sigma}$, $n_{c\gamma\sigma}$ and $n_{f\gamma\sigma}$, and their ${\mib k}$-integrated spectral weight or DOS, $\rho_{x\gamma\sigma}(\omega)$ ($x=W, f, c$) in the presence of the uniform magnetic field\cite{Newns87}.
For this purpose, we add the Zeeman term to the Hamiltonian, eq.~(\ref{hamiltonian}),
\begin{equation}
H_{\rm Zeeman}=-\sum_{{\mib k}\gamma\sigma}g_\gamma \sigma h(c^\dagger_{{\mib k}\gamma\sigma}c_{{\mib k}\gamma\sigma}+f^\dagger_{{\mib k}\gamma\sigma}f_{{\mib k}\gamma\sigma}),
\end{equation}
where we have used the same $g_\gamma$-factor both for conduction and $f$ electrons.
With use of the rectangular DOS for the conduction electron, i.e., $\rho_{c0}(\omega)=\theta(\omega+D)\theta(D-\omega)/2D$, $D$ being the half bandwidth , the spectral weights are given by
\begin{subequations}
\begin{align}
&\rho_{W\gamma\sigma}(\tilde{\omega})=\frac{1}{2D}\left(\frac{z_{\gamma\sigma}V}{\omega-\lambda_{\gamma\sigma}}\right)\theta_{\gamma\sigma}(\omega), \\
&\rho_{c\gamma\sigma}(\tilde{\omega})=\frac{1}{2D}\theta_{\gamma\sigma}(\omega), \\
&\rho_{f\gamma\sigma}(\tilde{\omega})=\frac{1}{2D}\left(\frac{z_{\gamma\sigma}V}{\omega-\lambda_{\gamma\sigma}}\right)^2\theta_{\gamma\sigma}(\omega),
\end{align}
\end{subequations}
where $\tilde{\omega}=\omega-\mu-g_\gamma\sigma h$.
The step function is defined as
\begin{equation}
\theta_{\gamma\sigma}(\omega)=
\sum_m\theta(\omega-E^{(m)}_{-\gamma\sigma})\theta(E^{(m)}_{+\gamma\sigma}-\omega),
\end{equation}
where the QP band edges are given by
\begin{equation}
E^{(\pm)}_{\pm\gamma\sigma}=E^{(\pm)}_{{\mib k}\gamma\sigma}(\epsilon_{\mib k}\to \pm D).
\end{equation}
The density is calculated by $n_{x\gamma\sigma}=\int_{-\infty}^{\tilde{\mu}_{\gamma\sigma}(h)}d\omega \rho_{x\gamma\sigma}(\tilde{\omega})$,
where $\tilde{\mu}_{\gamma\sigma}(h)=\mu+g_\gamma\sigma h$.
The explicit expressions depend on the position of the chemical potential:
\begin{enumerate}
\item $\tilde{\mu}_{\gamma\sigma}(h) < E^{(-)}_{-\gamma\sigma}$ \\
\begin{equation}
W_{\gamma\sigma}=n_{c\gamma\sigma}=n_{f\gamma\sigma}=0,
\end{equation}
\item $E^{(-)}_{-\gamma\sigma} < \tilde{\mu}_{\gamma\sigma}(h) < E^{(+)}_{-\gamma\sigma}$ \\
\begin{subequations}
\begin{align}
&W_{\gamma\sigma}=\frac{z_{\gamma\sigma}V}{2D}\ln\left(\frac{\lambda_{\gamma\sigma}-E^{(-)}_{0\gamma\sigma}(h)}{\lambda_{\gamma\sigma}-E^{(-)}_{-\gamma\sigma}}\right), \\
&n_{c\gamma\sigma}=\frac{1}{2D}\left(E^{(-)}_{0\gamma\sigma}(h)-E^{(-)}_{-\gamma\sigma}\right), \\
&n_{f\gamma\sigma}=\frac{q_{\gamma\sigma}V^2}{2D}\left(\frac{1}{\lambda_{\gamma\sigma}-E^{(-)}_{0\gamma\sigma}(h)}-\frac{1}{\lambda_{\gamma\sigma}-E^{(-)}_{-\gamma\sigma}}\right),
\end{align}
\end{subequations}
\item $E^{(+)}_{-\gamma\sigma} < \tilde{\mu}_{\gamma\sigma}(h)$ \\
\begin{subequations}
\begin{align}
&W_{\gamma\sigma}=\sum_m\frac{z_{\gamma\sigma}V}{2D}\ln\left(\frac{\lambda_{\gamma\sigma}-E^{(m)}_{0\gamma\sigma}(h)}{\lambda_{\gamma\sigma}-E^{(m)}_{-\gamma\sigma}}\right), \\
&n_{c\gamma\sigma}=\sum_m\frac{1}{2D}\left(E^{(m)}_{0\gamma\sigma}(h)-E^{(m)}_{-\gamma\sigma}\right), \\
&n_{f\gamma\sigma}=\sum_m\frac{q_{\gamma\sigma}V^2}{2D}\left(\frac{1}{\lambda_{\gamma\sigma}-E^{(m)}_{0\gamma\sigma}(h)}-\frac{1}{\lambda_{\gamma\sigma}-E^{(m)}_{-\gamma\sigma}}\right).
\end{align}
\end{subequations}
\end{enumerate}
Here we have defined the upper bound of the integral, $E^{(\pm)}_{0\gamma\sigma}(h)=\min(\tilde{\mu}_{\gamma\sigma}(h),E^{(\pm)}_{+\gamma\sigma})$.

\end{document}